\documentstyle[11pt,epsf]{article}

\newcommand{\be}{\begin{equation}}
\newcommand{\ee}{\end{equation}}
\newcommand{\bea}{\begin{array}}
\newcommand{\ea}{\end{array}}
\newcommand{\beqa}{\begin{eqnarray}}
\newcommand{\eeqa}{\end{eqnarray}}
\newcommand{\bean}{\begin{eqnarray*}}
\newcommand{\eean}{\end{eqnarray*}}

\def\up#1{\leavevmode \raise.16ex\hbox{#1}}

\begin{document}

\title{\hfill$\mbox{\small{
$\stackrel{\rm\textstyle DSF-1997/48} 
{\rm\textstyle quant-ph/9712028}$}}$\\[.6truecm]
Quantum singular oscillator as a model of two-ion trap:
an amplification of transition probabilities due to small time variations
of the binding potential}
\author{ 
V.V. Dodonov$^{a,b}$\thanks{e-mail: vdodonov@power.ufscar.br},\,
V. I. Man'ko$^{b,c}$\thanks{e-mail: manko@na.infn.it}\,
and L. Rosa$^{c,d}$\thanks{e-mail: rosa@na.infn.it } 
}
\maketitle

\thispagestyle{empty}

\begin{center}
$ ^{a)}$ Departamento de F\'{\i}sica, Universidade Federal de S\~ao Carlos,\\
Via Washington Luiz, km 235, 13565-905  S\~ao Carlos,  SP,  Brazil \\
$ ^{b)}$ Lebedev Physical Institute, Leninsky Prospect 53, 117924 Moscow,
Russia \\
$ ^{c)}$  Dipartimento di Scienze Fisiche, Universit\`a di Napoli,\\
Mostra d'Oltremare, Pad.19, I-80125, Napoli, Italy \\
$ ^{d)}$ INFN, Sezione di Napoli, Napoli, Italy 
\end{center}

\begin{abstract}
Following the paper by M. Combescure [Ann. Phys. (NY) {\bf 204}, 113
(1990)], we apply the quantum singular time dependent oscillator model to
describe the relative one dimensional motion of two ions in a trap. We
argue that the model can be justified for low energy excited states with
the quantum numbers $n\ll n_{max}\sim 100$, provided that the dimensionless
constant characterizing the strength of the repulsive potential is large
enough, $g_*\sim 10^5$. Time dependent Gaussian-like wave packets
generalizing odd coherent states of the harmonic oscillator, and excitation
number eigenstates are constructed. We show that the relative motion of the
ions, in contradistinction to its center of mass counterpart, is extremely
sensitive to the time dependence of the binding harmonic potential, since
the large value of $g_*$ results in a significant amplification of the
transition probabilities between energy eigenstate even for slow time
variations of the frequency. 
\end{abstract}


\section{Introduction}

A quantum system described by the singular oscillator Hamiltonian
\begin{equation}
\hat {H}=\frac {\hat {p}^2}{2\mu}+\frac 12\mu\omega^2(t)x^2+\frac g{
x^2}
\label{Ham}
\end{equation}
is one of rare examples admitting exact solutions of the Schr\"odinger
equation. In the case of a constant frequency $\omega$, the eigenfunctions
can be found in textbooks \cite{Land}. Generalizations of these eigenfunctions
to the case of a time dependent frequency were obtained for the first time
in \cite{Camiz}. Using these solutions, an exact propagator of the
Schr\"odinger equation with Hamiltonian (\ref{Ham}) was found
in \cite{DMM72}. An elegant method of calculating propagators of
time dependent quantum systems, including (\ref{Ham}), on the basis of
quantum operator integrals of motion, was developed in \cite{DMM75}.
The same propagator was obtained in the framework of the path integrals
approach in \cite{Inom} (a constant frequency) and \cite{Khan} (a time
dependent frequency). A detailed analysis of solutions found in
\cite{Camiz} was performed in Ref. \cite{DMM74},
where two families of
time dependent wave packets generalizing coherent states of the harmonic
oscillator \cite{gla63} were also constructed.
An asymmetrical singular nonstationary oscillator in two and three
dimensions was considered in \cite{DMM74b} (see also \cite{Haut} for the
stationary case).
A systematic study of wave packets and
coherent states in the centrifugal and other anharmonic potentials was given in
\cite{Niet}.

The exact solvability of the Schr\"odinger equation with Hamiltonian
(\ref{Ham}) is explained by the fact that this Hamiltonian is a linear
combination of three generators of the $su(1,1)$ algebra,
\be
\frac{p^2}{2\mu} +\frac{g}{x^2}, \quad x^2, \quad xp+px .
\label{su11}
\ee
The propagator and solutions of the Schr\"odinger equation with
a generic Hamiltonian, which is a linear combination of generators
(\ref{su11}) with arbitrary time dependent coefficients, were found
in \cite{CDM86} (see also \cite{Prants,Datt}, where other representations
of the $su(1,1)$ algebra were considered).
The universal invariants and generalized uncertainty
relations for this generic case were studied in \cite{univ}. A special
case of periodic time dependent coefficients, including the adiabatic
(Berry's phase) and delta-kicked limits, was
considered in \cite{Zhiv}. Recently, many results obtained in the cited papers
were rediscovered, see e.g. \cite{Ag,Maam,Kaush}.

A distinguished role of Hamiltonian (\ref{Ham})
is explained by the fact that, in a sense, it belongs to a boundary between
linear and nonlinear problems of classical and quantum mechanics. For this
reason, it was used in many applications. For example, it
served as an initial point in constructing interesting
exactly solvable models of interacting $N$-body systems \cite{Calog}.
It was used also for modelling diatomic \cite{CDM86} and polyatomic
\cite{Hart} molecules, and for the
evaluation of the accuracy of the quasiclassical approximation for
nonquadratic Hamiltonians \cite{167}.
Studying the limiting case $g= 0$ of the generalized coherent states of a
singular oscillator led to introducing a concept of {\it even and odd
coherent states\/} \cite{DMM74}, which are considered nowadays as simplest
models of quantum macroscopic superpositions --- the so called Schr\"odinger
cat states \cite{sch35}.

Even and odd coherent states of photons were
created in cavity experiments \cite{haroche}.
Another superposition of coherent states of the Schr\"odinger cat type,
similar to even and odd coherent states, can be created from an initial
coherent state in a nonlinear Kerr medium \cite{ys}.
This superposition has the same Poisson quantum number distribution function
as the usual coherent state, so it is ``more classical'' \cite{octavio}
than even and odd coherent states, which
have highly oscillating quantum number distribution functions (strong
oscillations of photon distribution functions, like in
squeezed states \cite{yuen,walls}, are believed to be
a characteristics of nonclassical states \cite{schleich}).
The photon statistics of parametrically excited even and odd coherent
states (which are superpositions of squeezed states) was discussed in
\cite{nikonov}.
A {\it quantum sling} mechanism of constructing squeezed and correlated
states from usual coherent states was suggested in \cite{hacyan}.

Unfortunately, photon nonclassical states are very
sensitive to the interaction with an environment, and decoherence
phenomena may destroy the specific interference pattern of the
Schr\"odinger cat superposition states very rapidly
(for example, an increase of temperature leads to a decrease of
oscillations of the photon distribution function in squeezed and
correlated states \cite{rosa}).
Much more stable even and odd coherent states can be
created for trapped ions (for example, in the Paul trap \cite{paul}),
as was demonstrated both theoretically \cite{vog1,gerry} and experimentally
\cite{wineland}.
Different aspects of quantum motion of ions in traps were discussed, e.g.
in \cite{octavio,glaconf,schrama}. In particular, various
 nonlinear modifications of coherent states \cite{sudphysscr}
 could be created for trapped ions \cite{vog2}.

Here we wish to show that the limiting case of very large values of the
dimensionless parameter $\mu g/\hbar^2 $ is also interesting,
because it may have some relation to the problem of the {\it relative\/}
motion of ions in electromagnetic traps.
Indeed, the ions in the Paul or Penning traps
move in time dependent {\it harmonic\/} potentials, and the stable
configurations correspond under certain conditions to an effectively
unidimensional motion along the trap axis \cite{Blum}. In such a case,
a relative motion of two ions can be described as the motion of a single
particle (with the reduced mass) in the Coulomb repulsive potential
$V(x)=e^2/x$.
Combescure \cite{Comb} proposed to replace the real Coulomb potential by
the singular oscillator potential $V_{ef}=g/x^2$. Using the solutions of
the Schr\"odinger equation in the same form as in \cite{Camiz,DMM74} and
considering the harmonic time dependence of the frequency $\omega(t)$, he
made some qualitative conclusions concerning the relative motion of the
ions in the Paul trap. However, he did not give any estimations which could
justify his model. The aim of our article is to provide such a
justification for large values of parameter $g$, and to consider specific
features of solutions in this special case, stressing qualitative
differences between the relative and center of mass motions. We shall
demonstrate that the transition probabilities between energy eigenstates of
the relative motion are significantly amplified in comparison with that of
the center of mass motion, even when time variations of the frequency are
very small. This result may be interpreted as an indication to a high
sensitivity of the relative ion motion to small time variations of the
binding trap potential, which could result in some kind of a chaotic
behaviour in the case of the real Coulomb interaction potential
\cite{Blum,Nat,Moor}. Note that an irregular (although not chaotic, in the
strict sense of this notion) behaviour of time dependent {\it linear
quantum systems\/} described by $su(1,1)$ invariant Hamiltonians with
(quasi)periodic $\delta$-kicked coefficients was studied in
\cite{Ger89,Ger90} 
(see also \cite{Dod96}).

\section{Justification of the model}

Let us compare two potentials ($\mu$ is the reduced mass of two ions),
\be
V(x)=\frac12 \mu\omega^2 x^2 +\frac{e^2}{x},
\label{V}
\ee
\be
V_{g}(x)=\frac12 \mu\omega_{g}^2 x^2 +\frac{g}{x^2}.
\label{V-g}
\ee
If the energy of a particle is not too large (as in real traps, where
ions are cooled), then it performs small oscillations near the equilibrium
coordinates satisfying the equations $V'(x)=0$ and $V'_{g}(x)=0$,
namely,
\[
x_e=\left(\frac{e^2}{\mu\omega^2}\right)^{1/3}, \quad
x_g=\left(\frac{2g}{\mu\omega_g^2}\right)^{1/4}.
\]
The frequencies of oscillations are determined by the second derivatives
of the potentials in the equilibrium points,
\[
\Omega_{e,g}=\sqrt{V''\left(x_{e,g}\right)/\mu}, \quad
\Omega_e=\sqrt3\omega, \quad \Omega_g=2\omega_g.
\]
Demanding the equilibrium positions and the effective frequencies
of the real potential, (\ref{V}), and its simulation, (\ref{V-g}),
to coincide , we obtain the following parameters of the effective
potential describing the relative ion motion:
\be
\omega_{g}=\frac{\sqrt3}{2}\omega, \quad
g=\frac38\left(\frac{e^8}{\mu\omega^2}\right)^{1/3}.
\label{g-e}
\ee
The minimum is twice less than that of the real potential,
\[
V\left(x_e\right)=\frac32\mu\omega^2 x_e^2, \quad
V_g\left(x_g\right)=\mu\omega_g^2 x_g^2=\frac12 V\left(x_e\right),
\]
but this shift seems not important.

The solutions of the Schr\"odinger equation depend on the dimensionless
parameter
\be
g*= 2\mu g/\hbar^2.
\label{g*}
\ee
Taking the $x$-dependent part of the trap potential energy as
$V_{t}(x)=eUx^2/(2 L^2)$,
$U$ being the voltage applied to the cap electrodes separated by
the distance $2L$, we have $\omega^2= eU/(2\mu L^2)$ (since $\mu=m_i/2$,
where $m_i$ is the ion mass). Therefore,
\be
g_*=
\frac34\left(\frac{\mu^2 e^8}{\hbar^6\omega^2}\right)^{1/3}=
\frac{3\mu}{4m}\left[\frac{4\mbox{Ry}}{eU}\left(\frac{L}{a_B}\right)^2
\right]^{1/3},
\label{g*est}
\ee
where $m$ is the electron mass, $\mbox{Ry}$ is the Rydberg constant,
and $a_B$ is Bohr's radius. For typical trap parameters, $\mu/m\sim 10^5$,
$U\sim 100$~V, $L\sim 1$~mm, we have $g_*\sim 10^{10}$.

A specific feature of the singular oscillator potential is the
{\it equidistant\/} character of the energy spectrum. Since the real
interaction potential does not possess such a property, the replacement of
(\ref{V}) by (\ref{V-g}) can be justified, provided that the corrections
to the energy spectrum due to the anharmonicity of the potential
(\ref{V}) are small, comparing with the effective
 oscillator energy difference $\hbar\Omega_e=\sqrt3\hbar\omega$.
The cubic
anharmonic correction to the potential (\ref{V}) in the vicinity of the
equilibrium point is $-\left(e^2/x_e^4\right)(\delta x)^3$, $\delta x$ being
a small deviation from the equilibrium position $x_e$. Assuming
$\mu\Omega_e^2(\delta x)^2=n\hbar\Omega_e$ for the $n$th excited
effective harmonic oscillator energy level, one can verify that the
anharmonic corrections can be neglected under the condition
\be
n\ll n_{max}\sim 3\left(\frac{\mu e^4}{\hbar^3\omega}\right)^{1/9}
\sim 3g_*^{1/6}.
\label{n-max}
\ee
For the same trap parameters as above, $n_{max}\approx 100$. Consequently,
the singular oscillator model can be applied for describing the low energy
excited states of the relative ion motion.

\section{Time dependent solutions for a large repulsive constant}

A complete set of orthonormalized solutions to the Schr\"odinger equation
with Hamiltonian (\ref{Ham}) can be chosen in the form \cite{Camiz,DMM74}
\be
\Psi_n(x,t)=\left[ 2\left( \frac{\mu}{\hbar\epsilon^2} \right)^{d+1}
\frac{n!}{\Gamma(d+n+1)}\right]^{\frac{1}{2}}
\left(\frac{\epsilon^*}{\epsilon}\right)^{n} x^{d+\frac{1}{2}}
\exp{\left(\frac{i \mu\dot\epsilon}{2\hbar\epsilon}x^2 \right)}
L^d_n\left( \frac{\mu x^2}{\hbar|\epsilon|^2}  \right),
\label{sol-Lag}
\ee
where
$d=\frac{1}{2}\sqrt{1+4g_*}$, and $L^d_n(z)$ is an associated Laguerre
polynomial. (Hereafter we write $\omega$ instead of $\omega_g$, assuming
that the frequency rescaling has been done.)
The time dependent complex function $\epsilon(t)$
is a special solution to the classical harmonic oscillator equation,
\be
\ddot\epsilon+\omega^2(t)\epsilon=0, \label{epsil}
\ee
satisfying the normalization condition
\be
\dot\epsilon\epsilon^*-\dot\epsilon^*\epsilon=2i. \label{normep}
\ee
If $\omega=const$, the solutions (\ref{sol-Lag}) go to the energy eigenstates,
provided that function $\epsilon(t)$ is chosen as
$\epsilon(t)=\omega^{-1/2}\exp\left(i\omega t\right)$.

Expression (\ref{sol-Lag}) is written for $x>0$. Its continuation to the
negative semiaxis (since $x$ is the relative coordinate of two ions along
the trap symmetry axis, it can be both positive and negative) depends on
the symmetry properties of the two-particle wavefunction determined by the
kind of statistics (Bose or Fermi, if the ions are identical), so that
one should write $\Psi(x)=\pm\Psi(-x)$ for $x<0$. There are no problems
with the continuity of the solutions or their derivatives at $x=0$,
provided that $d>1/2$ (i.e. $g>0$). Extending the wave function to the
whole axis, one must only change the normalization by the evident factor
$\sqrt2$. Authors of the recent paper \cite{Feng} claimed that there is
a nontrivial influence of the parameter $d$ on the possibility of
existence of the bound states of two ions in the Paul trap, depending on
the type of quantum statistics. However, this conclusion was a consequence of
incorrect manipulations with the expressions for the wave function
at $x>0$ and $x<0$, which resulted in the coordinate independent
factor $1\pm \cos(d+1/2)\pi$ in the wave function.
This factor turns into zero for specific values
of parameter $d$, which was interpreted by the authors of \cite{Feng} as
a disappearance of bound states. They forgot, however, to normalize the
wave function. If they did this, then the incorrect factor would disappear,
not the bound states.

Solution (\ref{sol-Lag}) is an eigenfunction of the time dependent
integral of motion (a generalization of the harmonic oscillator number
operator)
\be
\hat {B}=\frac 1{\hbar}\left[|\epsilon|^2\left(\frac {\hat {p}^
2}{2\mu}+\frac g{x^2}\right)+\frac {\mu}{2}|\dot\epsilon|^2
 x^2-\frac 12\mbox{Re}\left(\dot\epsilon\epsilon^*\right)
 \left(\hat x\hat p+\hat p\hat x\right)\right],\label{int-B}
\ee
with an eigenvalue $2n+d+1$.
There exists another integral of motion \cite{DMM74},
\be
\hat {A}=\frac {1}{2\mu\hbar}\left[\left(\epsilon\hat p-\dot
\epsilon \mu\hat x\right)^2+2\epsilon^2\frac {\mu g}{x^2}\right],
\label{int-A}
\ee
which is a generalization of the square of the harmonic oscillator
annihilation operator $\hat a^2$. A normalized eigenstate of operator
(\ref{int-A}),
$\hat {A}|\alpha \rangle = \alpha^2|\alpha \rangle$ ($\alpha$ being an
arbitrary complex number), reads \cite{DMM74}
\beqa
\langle x|\alpha\rangle &=&
\frac{|\alpha|^d}{\sqrt{2^d I_d\left(|\alpha |^2\right)}}
\sum_{n=0}^{\infty}
\frac{\alpha^{2n}}{\sqrt{2^{2n} n! \Gamma(n+d+1)}}
\Psi_n(x,t)
\label{alpha-n}
\\
& =&\left[\frac {2\mu x}{\hbar\epsilon^2
I_d\left(|\alpha |^2\right)}\right]^{1/2}
\left(\frac{|\alpha|}{\alpha}\right)^d
\exp\left(\frac {i\mu\dot{\epsilon}}{
2\hbar\epsilon}x^2+\frac {\epsilon^{*}}{2\epsilon}\alpha^
2\right)J_d\left(\frac {x\alpha}{\epsilon}\sqrt {\frac {2\mu}{\hbar}}
\right),\label{sol-Bes}
\eeqa
where $J_d(x)$ and $I_d(x)$ are the Bessel function and the
modified Bessel function, respectively.
Similar (time independent) eigenstates of the lowering generator of an
abstract $su(1,1)$ algebra in another specific representation
were considered by Barut and Girardello \cite{BG}.
Various generalizations and concrete realizations of
these states were discussed, e.g. in \cite{Buz,Sat,Trif,Brif},
where many references to other publications can be found.

Taking $d=1/2$ (i.e. $g=0$) and replacing the reduced mass, $\mu$, by the
total mass of two ions, $M$, we arrive at the expressions for the wave
functions describing the {\it odd\/} states of the center of mass motion
in the harmonic potential with frequency $\omega$ (remember, however, that
the frequency of small oscillations of the relative distance between ions
is $\sqrt3$ times higher than the frequency of the center of mass motion).
The {\it even\/} wave functions of the center of mass motion
are obtained by means of a formal substitution $d=-1/2$.

The time dependences of the mean values of the operators (\ref{su11})
in any state are as follows,
\beqa
\langle x^2\rangle &=&
\frac{\hbar}{m}\left[|\epsilon|^2
\langle \hat B\rangle - \mbox{Re}\left(\epsilon^2
\langle \hat A^{\dag}\rangle\right)\right], \label{mean-x2}\\
\left\langle \frac {\hat {p}^2}{2\mu}+\frac g{x^2}\right\rangle &=&
\frac{\hbar}{2}\left[|\dot\epsilon|^2
\langle \hat B\rangle - \mbox{Re}\left(\dot\epsilon^2
\langle \hat A^{\dag}\rangle\right)\right], \label{mean-p2}\\
 \left\langle\hat x\hat p+\hat p\hat x\right\rangle &=&
2\hbar\left[\mbox{Re}\left(\dot\epsilon\epsilon^*\right)
\langle \hat B\rangle - \mbox{Re}\left(\dot\epsilon \epsilon
\langle \hat A^{\dag}\rangle\right)\right].
\label{mean-px}
\eeqa
In particular, the mean value of the operator $\hat{B}$ in the state
(\ref{sol-Bes}) equals \cite{DMM74}
\be
\langle\alpha|\hat{B}|\alpha\rangle= 1 +|\alpha|^2
\frac{I'_d\left(|\alpha|^2\right) }{I_d\left(|\alpha|^2\right)}.
\label{aBa}
\ee

Formulas (\ref{sol-Bes}) and (\ref{aBa}) can be simplified
for large values of coefficient $d$. Taking into account the asymptotics of
the Bessel functions of a large index \cite{Bate},
\be
\sqrt{2\pi}I_d(z)\approx \left(d^2+z^2\right)^{-1/4}
\exp\left[\sqrt{d^2+z^2}
+d \ln\left(\frac{z}{z+\sqrt{d^2+z^2}}\right)\right],
\quad d\gg 1, \quad d>z.
\label{as-Bes}
\ee
and calculating the power series
expansion of the exponential argument in (\ref{as-Bes}) up to
$z^4$ terms, we obtain
\be
\langle\alpha|\hat{B}|\alpha\rangle= 1 +d +\frac{|\alpha|^4}{2d}.
\label{aBa-as}
\ee
Consequently, the energy difference between the states $|\alpha\rangle$
and $|0\rangle$ becomes noticeable only for $|\alpha|^4\sim d$.
In such a case, due to Eq. (\ref{mean-x2}), the mean value
$\langle x^2\rangle$ oscillates in the vicinity of the average position
$\langle x_0^2\rangle = (\hbar/m)|\epsilon|^2 \langle \hat B\rangle \sim 2$,
the amplitude of oscillations being of an order of $\sqrt{d}\ll d$.
For this reason, we can simplify formula (\ref{sol-Bes}), replacing
the Bessel function by its asymptotical form following from (\ref{as-Bes})
and expanding the arising exponential argument up to the terms of the
order of $|\alpha|^4/d$ or $x^2|\alpha|^2/d$. After some algebra, we
obtain the expression
\beqa
\langle x|\alpha\rangle &=&(2\pi d)^{-1/4}\left(\frac{2\mu x}
{\hbar\epsilon^2}\right)^{1/2}\left(\frac{\epsilon^*}{\epsilon}\right)^{d/2}
\exp\left[-\frac {\mu^2 y^2}{4\hbar^2|\epsilon|^4 d}
+\frac{i\mu}{2\hbar}\mbox{Re}\left(\frac{\dot\epsilon}{\epsilon}\right)y
-\frac{|\alpha|^4}{8d}\right. \nonumber\\
&&-\left. \frac{\alpha^4\epsilon^{*2}}{8d\epsilon^{2}} -
\frac{\alpha^2 y}{2\epsilon^{2}d} \right],
\label{sol-Bes-as}
\eeqa
where
\be
y(x,t)=x^2 - \frac{\hbar }{\mu}|\epsilon|^2 (d+1).
\label{def-y}
\ee
In the right hand side of Eq. (\ref{sol-Bes-as}) we recognize the generating
function of the Hermite polynomials \cite{Bate},
\[
\exp\left( -z^2 + 2zx\right)=\sum_{n=0}^{\infty} \frac{z^n}{n!}H_n(x).
\]
Comparing it with the expansion (\ref{alpha-n}) and using
Stirling's formula, we arrive at the following asymptotical expression:
\be
\Psi_n(x,t)=
(2\pi d)^{-1/4}\left(\frac{2\mu x}
{\hbar\epsilon^2 2^n n!}\right)^{1/2}
\left(\frac{\epsilon^*}{\epsilon}\right)^{n+d/2}
\exp\left[-\frac {\mu^2 y^2}{4\hbar^2|\epsilon|^4 d}
+\frac{i\mu}{2\hbar}\mbox{Re}\left(\frac{\dot\epsilon}{\epsilon}\right)y
\right]
H_n\left( -\frac{\mu y}{\sqrt{2d}\hbar |\epsilon|^2}\right).
\label{n-herm}
\ee
The probability density in the state (\ref{sol-Bes-as})
is Gaussian, but with respect to variable $x^2$:
\be
|\langle x|\alpha\rangle|^2 =(2\pi d)^{-1/2}\frac{2\mu x}
{\hbar|\epsilon|^2}
\exp\left[-\frac {\mu^2 \left(x^2- \langle\alpha| x^2|\alpha\rangle
\right)^2}{2\hbar^2|\epsilon|^4 d}\right],
\label{Gauss}
\ee
\be
\langle\alpha| x^2|\alpha\rangle=
\frac{\hbar}{m}\left[|\epsilon|^2
\left(1 +d +\frac{|\alpha|^4}{2d} \right)
-\mbox{Re}\left(\epsilon^{*2}\alpha^2\right)\right].
\label{mean-x2-a}
\ee
Exact solutions to the Schr\"odinger equation in the form of the
power-Gaussian wave packets were constructed in \cite{Camiz,DMM74}.
From the modern point of view, these packets can be considered as
generalizations of the {\it odd squeezed states\/}
(in the context of an abstract $su(1,1)$-invariant Hamiltonian similar
states were considered, e.g. in \cite{Onof}; recent results and
other references can be found in \cite{Trif,Brif}):
\beqa
|z\rangle &=& \left(1-|z|^2\right)^{\frac{d+1}{2}}
\exp\left(\frac{z}{2}\hat{A}^{\dag}\right)|0\rangle  \label{z-A+}\\
&=& \left[\frac{2\mu x}{\hbar\Gamma(d+1)}\right]^{1/2}
\left(\frac{\mu x^2}{\hbar}\right)^{d/2}
\left(\frac{\sqrt{1-|z|^2}}{\epsilon - z\epsilon^*}\right)^{d+1}
\exp\left[\frac{i\mu }{2\hbar}\frac{\dot\epsilon - z\dot\epsilon^*}
{\epsilon - z\epsilon^*}x^2\right].
\label{z-Gauss}
\eeqa
For not too large values of $d$, the states (\ref{sol-Bes}) and
(\ref{z-Gauss}) are obviously different. However, for
$d\gg 1$ and moderate values of $x$, $\alpha$, and $z$, the function
$|\langle x|z\rangle|^2$ is given by the same expression (\ref{Gauss}).
The only difference is in the formula for the mean value of $x^2$:
\be
\langle z| x^2|z\rangle=
(d+1)\frac{\hbar}{m}
\frac{|\epsilon - z\epsilon^*|^2}{1-|z|^2}
\approx (d+1)\frac{\hbar}{m}
\left[|\epsilon|^2 \left(1 +2|z|^2 \right)
-\mbox{Re}\left(\epsilon^{2}z^*\right)\right].
\label{mean-x2-z}
\ee
For moderately excited states, parameter $z$ must be small,
since \cite{DMM74}
\[
\langle z| \hat{B}|z\rangle= (d+1)\frac{1+|z|^2}{1-|z|^2}.
\]
Note that the relative width of the distributions given by Eqs.
(\ref{n-herm}) and (\ref{Gauss}) practically does not depend on time,
$\Delta y/y_{max}\sim\left(|\epsilon|^2\sqrt{d}\right)/
\left(|\epsilon|^2 d\right)\sim 1/\sqrt{d}$, while the position of the
center of the distribution, $y_{max}$, may vary essentially even with
small variations of function $|\epsilon|^2$, due to the large value of $d$.
As a consequence, the covering between the packets taken at different moments
of time may be very small even for weak variations of the frequency
$\omega(t)$, resulting in the effect of the amplification of
transition probabilities, discussed in the next section.

\begin{figure}[]
\epsfxsize=14cm
\epsffile{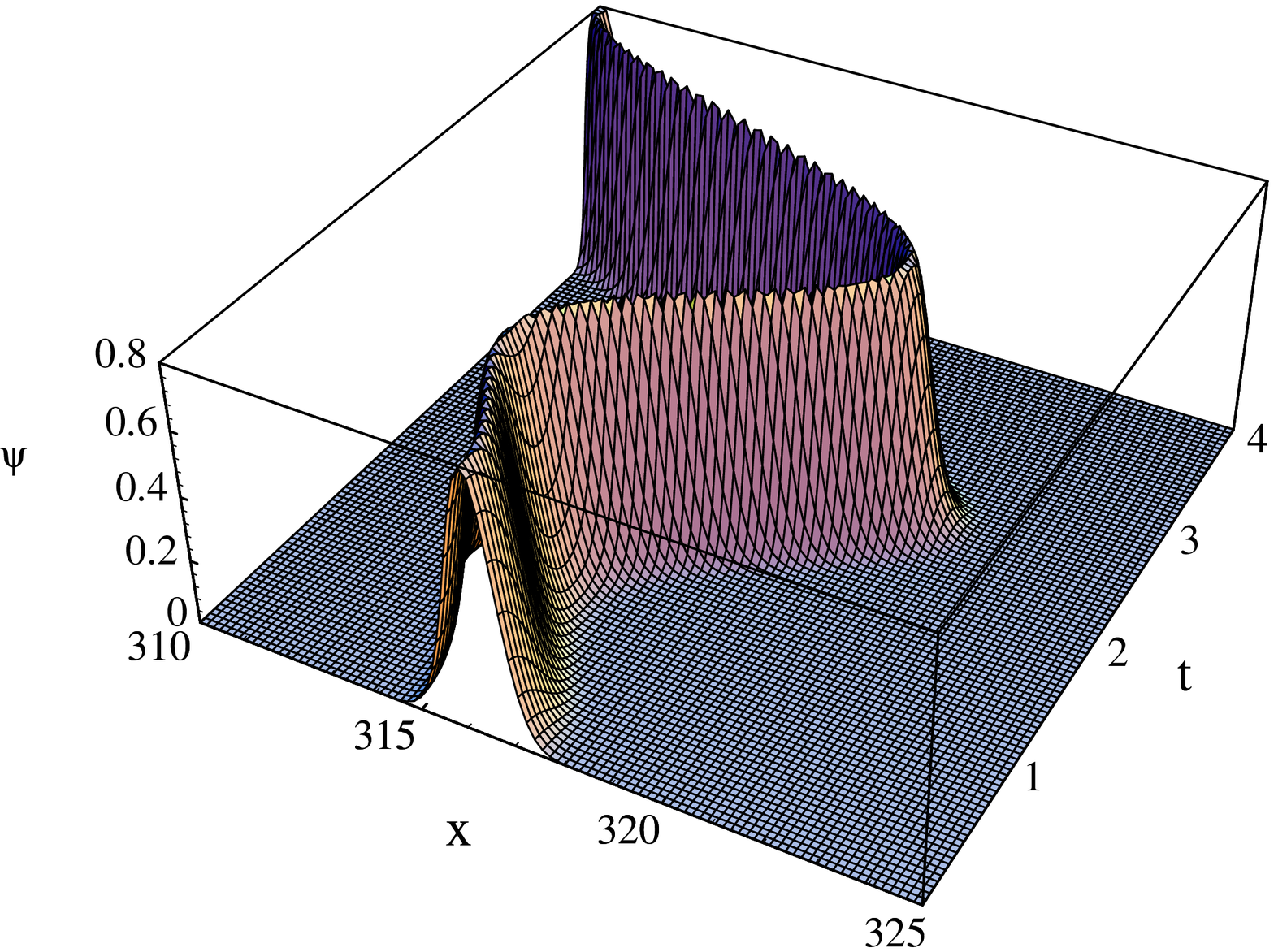}
\caption[]{ $|\Psi_0(x,t)|^2$ for $d=10^5$ and $k=0.02$ ($x$ and 
$t$ are dimensionless).}
\label{f:psi0}
\end{figure}

\begin{figure}[]
\epsfxsize=14cm
\epsffile{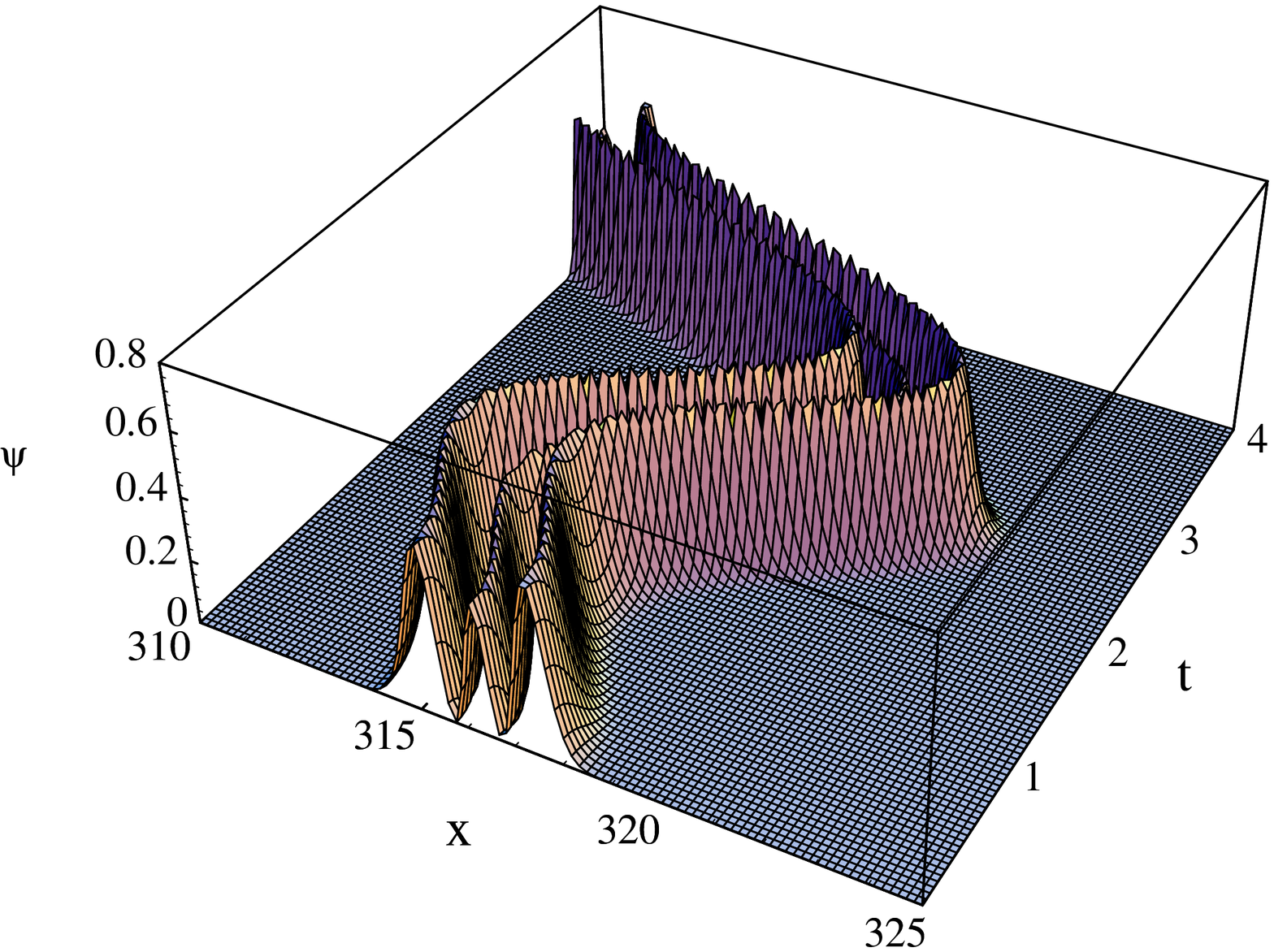}
\caption[]{ $|\Psi_2(x,t)|^2$ for $d=10^5$ and $k=0.02$ ($x$ and 
$t$ are dimensionless).}
\label{f:psi2}
\end{figure}

In Fig.\ref{f:psi0} and \ref{f:psi2}
we plot the probability densities for the states $\Psi_0(x,t)$ and 
$\Psi_2(x,t)$ (\ref{n-herm})
in the parametric resonance case (using dimensionless units)
\be
\omega^2(t)=\frac{1+k\cos{2t} }{1+k},~~k\ll1,
\label{par-res}
\ee
when an explicit analytical (approximate) expression for
$\epsilon(t)$ reads \cite{nikonov},
\be
\epsilon(t) = \cosh{\frac{kt}{4} }e^{it}-i\sinh{\frac{kt}{4} }e^{-it}.
\label{eps-k}
\ee
We assume $d=10^5$ and $k=0.02$.

\section{Amplification of transition probabilities}

Suppose that the frequency $\omega(t)$ assumes constant values, $\omega_i$ and
$\omega_f$, in the remote past and future, respectively
(such a situation corresponds to the Penning trap, when the voltage between
electrodes exhibits some time variations), and that the
singular oscillator was in the $n$th stationary state at $t\to-\infty$.
This state is given by Eq. (\ref{sol-Lag}) with
$\epsilon_i(t)=\omega_i^{-1/2}\exp\left(i\omega_i t\right)$.
The asymptotical form of the solution to Eq. (\ref{epsil}) at $t\to\infty$
reads
\[
\epsilon(t\to\infty)=\omega_f^{-1/2}\left[\xi\exp\left(i\omega_f t\right)
-\eta\exp\left(-i\omega_f t\right)\right],
\]
$\xi$ and $\eta$ being constant complex coefficients satisfying
the restriction $|\xi|^2-|\eta|^2=1$, following from (\ref{normep}).
The transition probability from the initial $n$th energy eigenstate to the
final $m$th one (corresponding to frequency $\omega_f$) was found in
\cite{DMM74}:
\beqa
&&W_n^m=\frac{\mu_{mn}!\Gamma\left(\nu_{mn}+d+1\right)}
{\nu_{mn}!\Gamma\left(\mu_{mn}+d+1\right)}r^{|m-n|}(1-r)^{d+1}
\left[P_{\mu_{mn}}^{(|m-n|,d)}(1-2r)\right]^2\label{probJac}\\
&&=\frac{\nu_{mn}!\Gamma\left(\nu_{mn}+d+1\right)r^{|m-n|}(1-r)^{d+1}}
{\mu_{mn}!\Gamma\left(\mu_{mn}+d+1\right)
[|m-n|!]^2}\left[F\left(-\mu_{mn},\nu_{mn}+d+1;|m-n|+1;r\right)\right]^2.
\label{prob-hyp}
\eeqa
Here $\mu_{mn}\equiv\mbox{min}(m,n)$, $\nu_{mn}\equiv\mbox{max}(m,n)$,
$\Gamma(z)$ is the Gamma function,
$P_n^{(\alpha,\beta)}(z)$ is the Jacobi polynomial, $F(a,b;c;z)$ is the
Gauss hypergeometric function \cite{Bate}, and $r\equiv|\eta/\xi|^2$ is
a reflection coefficient from an effective potential barrier determined by
the time dependence of $\omega^2(t)$.
In particular, the ground state excitation probability equals
\be
W_0^m=
\frac{\Gamma\left(m+d+1\right)}
{m!\Gamma\left(d+1\right)}r^m(1-r)^{d+1}.
\label{P0m}
\ee

In the special cases of $d=\pm 1/2$ (describing the center of mass motion
in the pure harmonic time dependent potential), one can use the
quadratic transformation of the Gauss hypergeometric function \cite{Bate},
\[
F(a,b;a+b+1/2;4z(1-z))=F(2a,2b;a+b+1/2;z),
\]
to reproduce the known formula for the transition probabilities between
the energy levels of the harmonic oscillator with a time dependent
frequency \cite{MM70} (see also \cite{183} for details and the bibliography),
\be
W_k^j=\frac{k!j!r^{|k-j|/2}\sqrt{1-r}}{2^{|k-j|}\left([(k+j)/2]!\right)^2}
\left[P_{\mu_{kj}}^{(|k-j|/2\,,\,|k-j|/2)}\left(\sqrt{1-r}\right)\right]^2.
\label{prob-osc}
\ee
Here $k$ and $j$ must have the same parity.

Now we have to take into account that
coefficient $d$ is very large, $d\approx\sqrt{g_*}\sim 10^5$,
in the case of the relative ion motion,
while quantum
numbers $m$ and $n$ must be much less than $100$. In such a case,
using an approximation $(b)_k\equiv b(b+1)\ldots(b+k-1)\approx b^k$, $b\gg 1$,
we can reduce the Gauss hypergeometric function
\[
F(a,b;c;z)\equiv \sum_{k=0}^{\infty}\frac{(a)_k (b)_k}{(c)_k k!}z^k
\]
to the confluent hypergeometric function
\[
\Phi(a;c;bz)\equiv \sum_{k=0}^{\infty}\frac{(a)_k }{(c)_k k!}(bz)^k.
\]
With the same accuracy, we can write $\Gamma(d+n)\approx \Gamma(d)d^n$.
Then Eq. (\ref{prob-hyp}), together with the relation between the confluent
hypergeometric function and the Laguerre polynomial \cite{Bate},
\[
\Phi(-n;c+1;z)\equiv \frac{n!}{(c+1)_n}L_n^c(z),
\]
lead to a simplified form of the transition probability for $d\gg 1$:
\be
W_n^m=\frac{\mu_{mn}!}{\nu_{mn}!}(rd)^{|m-n|} (1-r)^{d+1}
\left[L_{\mu_{mn}}^{|m-n|}(rd)\right]^2,
\label{prob-gbig}
\ee
\be
W_0^m=\frac{(rd)^m}{m!} (1-r)^{d+1}.
\label{P0m-gbig}
\ee

\begin{figure}[]
\epsfxsize=14cm
\epsffile{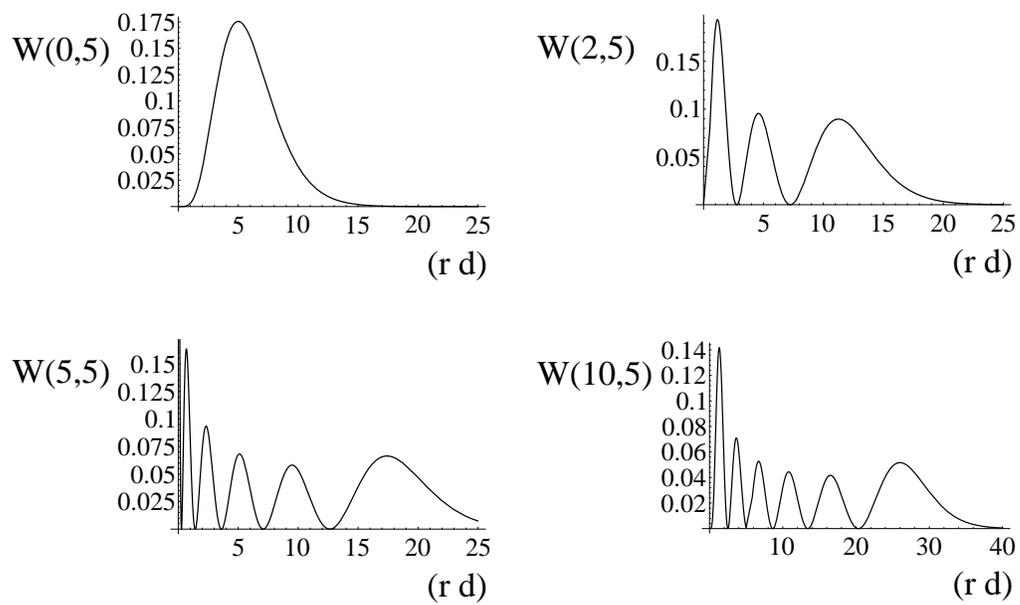}
\caption[]{ $W_5^m(r d)$ for $m=0,2,5,10$ ( $r$ and $d$ are 
dimensionless).}
\label{f:quad}
\end{figure}

\begin{figure}[]
\epsfxsize=14cm
\epsffile{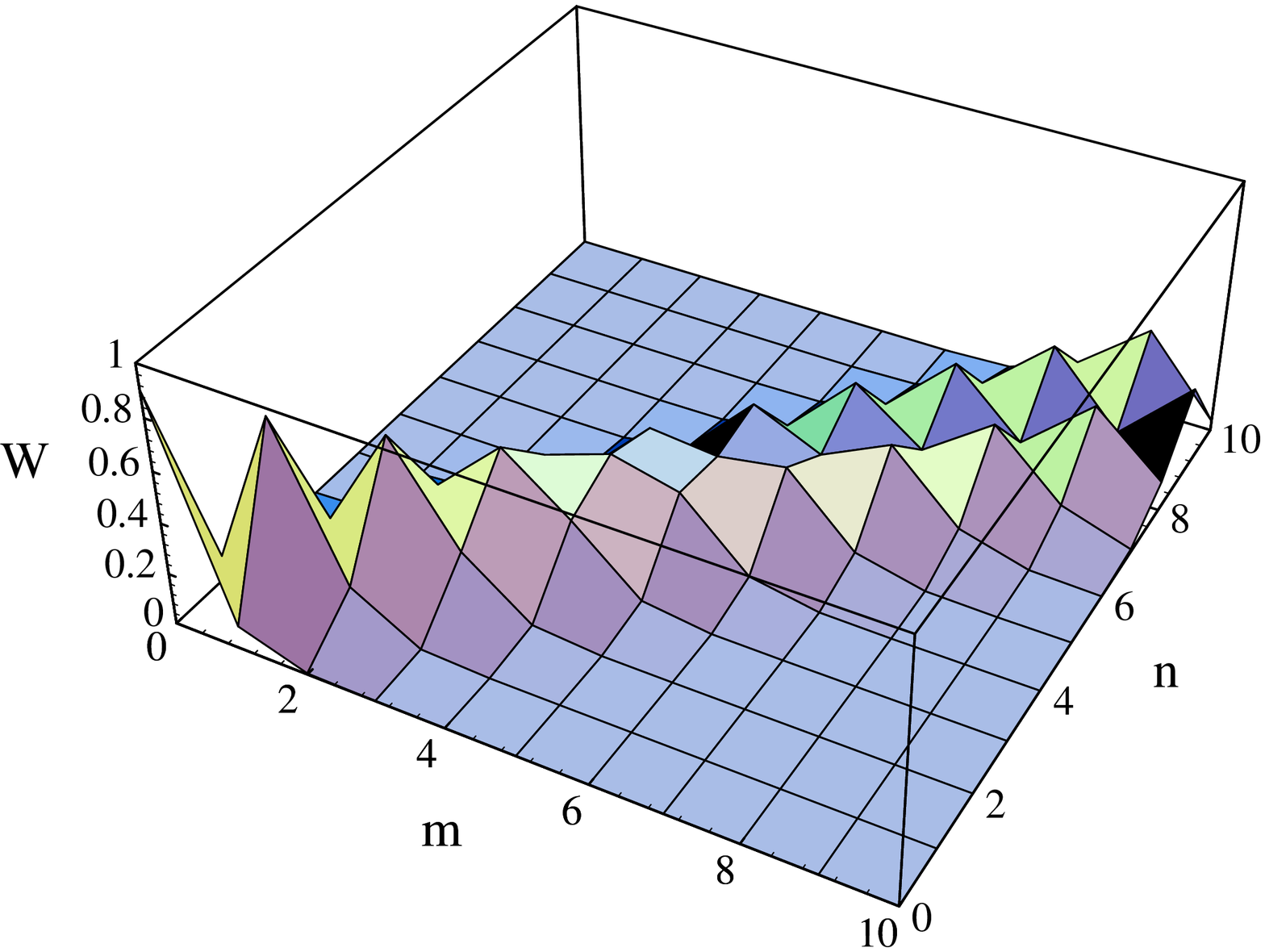}
\caption[]{ $W_n^m$ for $d=10^5$ and $r=10^{-6}$ }
\label{f:2wmnr6}
\end{figure}

\begin{figure}[]
\epsfxsize=14cm
\epsffile{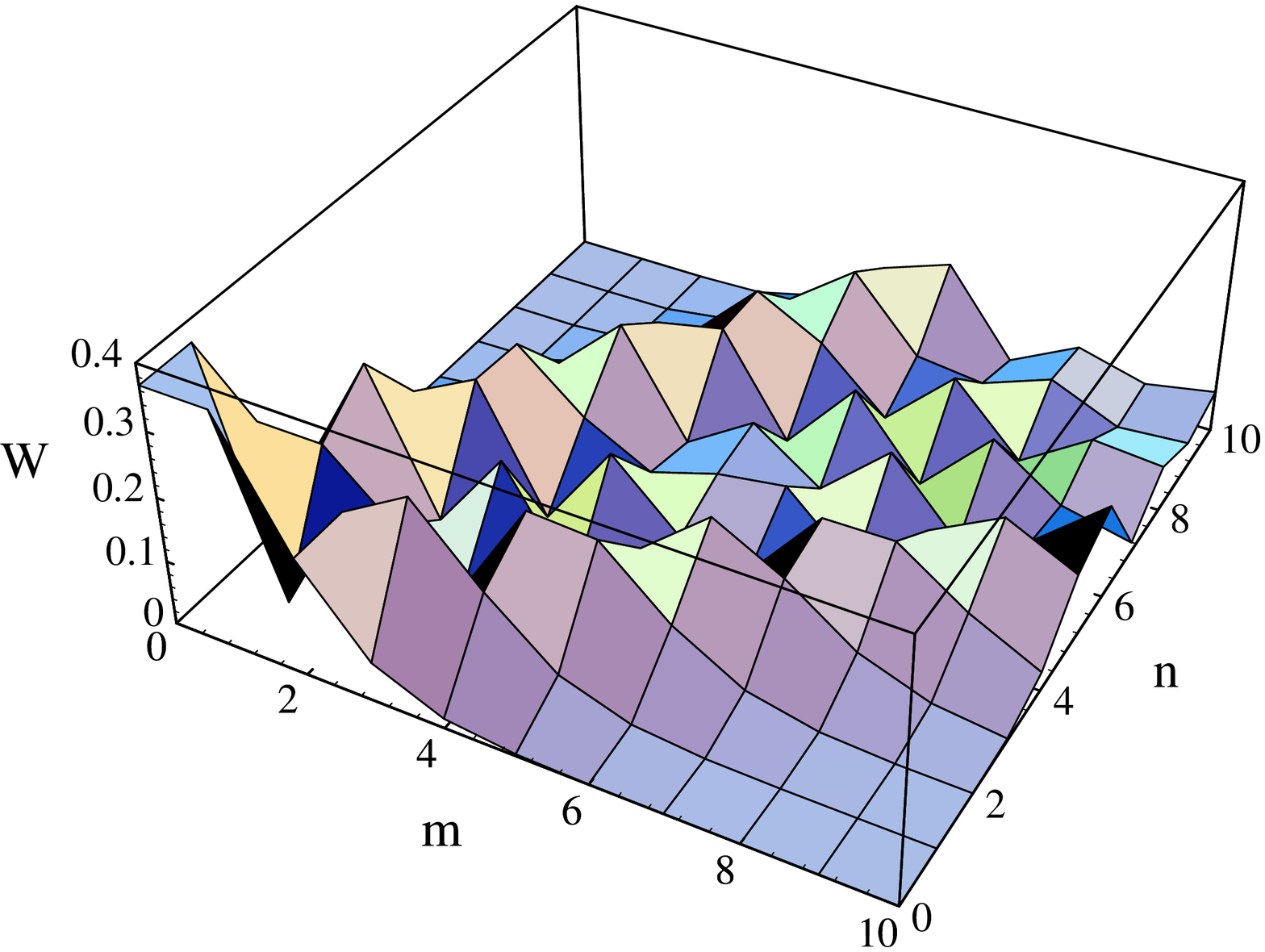}
\caption[]{ $W_n^m$ for $d=10^5$ and $r=10^{-5}$ }
\label{f:2wmnr5}
\end{figure}

\begin{figure}[]
\epsfxsize=14cm
\epsffile{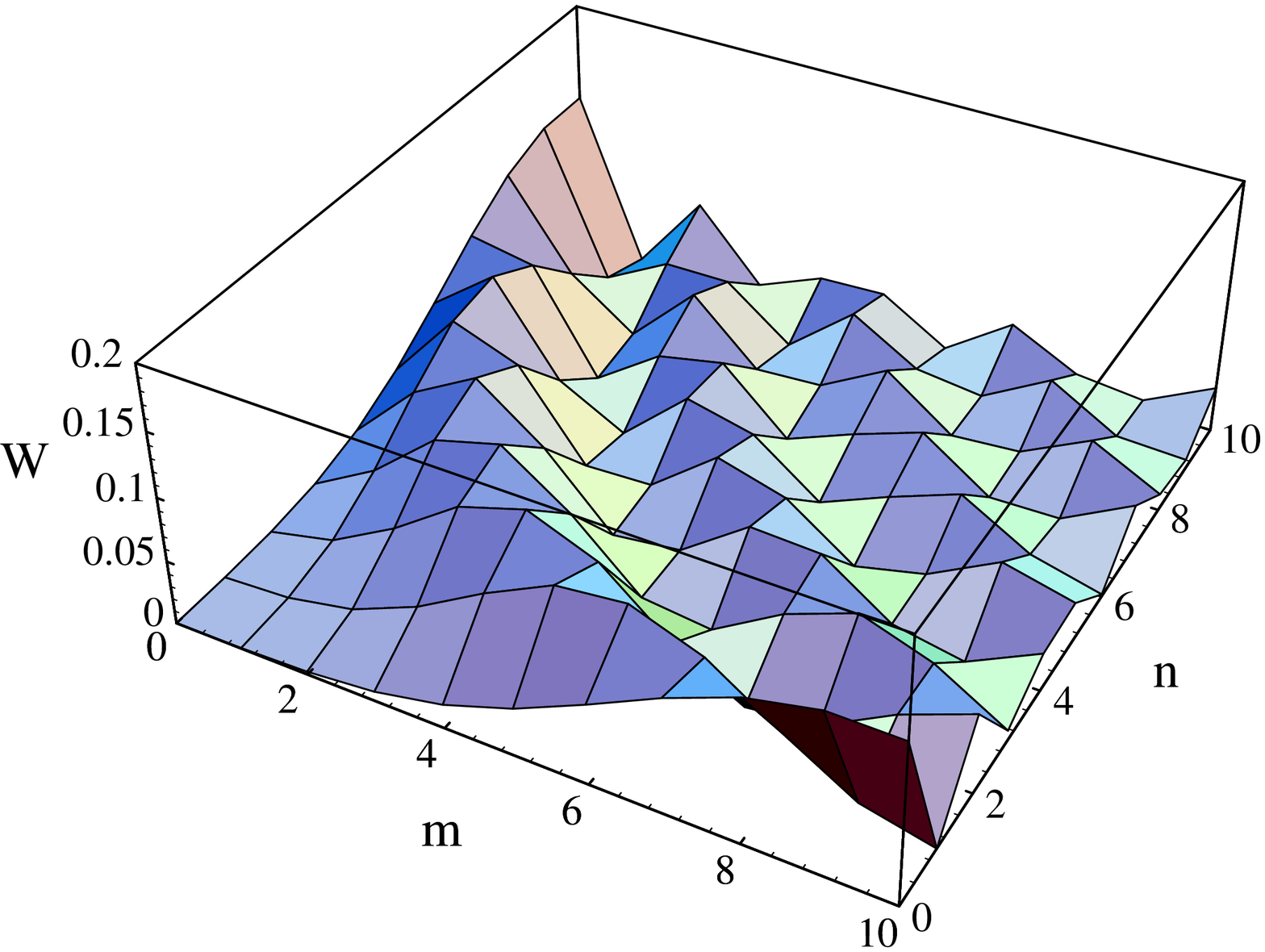}
\caption[]{ $W_n^m$ for $d=10^5$ and $r=10^{-4}$ }
\label{f:2wmnr4}
\end{figure}

If the frequency variation is small, for instance, adiabatic, then
the effective reflection coefficient $r$ is small too (excluding the case
of the parametric resonance). In this case, the Taylor expansion of
formula (\ref{prob-hyp}) yields
\be
W_n^m=
\frac{\nu_{mn}\Gamma\left(\nu_{mn}+d+1\right)}
{\mu_{mn}\Gamma\left(\mu_{mn}+d+1\right)}\frac{r^{|m-n|}}
{[|m-n|!]^2}\left[ 1- \frac{2mn +(d+1)(m+n+1)}{|m-n|+1}r +\cdots\right].
\label{prob-ad}
\ee
Consequently, if $r\ll 1$, and quantum numbers $m,n$ are not too large,
the transition probabilities of the center of mass motion ($d=\pm 1/2$)
are suppressed by the factor $r^{|m-n|}$. However, this may not be true
for the relative motion, since the reflection coefficient is effectively
amplified by the large factor $d$. This is seen distinctly from the
special form of formula (\ref{prob-gbig}), which is valid for $r^2d\ll 1$:
\be
W_n^m=\frac{\mu_{mn}!}{\nu_{mn}!}(rd)^{|m-n|} e^{-rd}
\left[L_{\mu_{mn}}^{|m-n|}(rd)\right]^2.
\label{prob-gbigg}
\ee
Fig.\ref{f:quad} gives the dependence of $W_n^m$ on
the product $rd$ for $n=5$ and $m=0,2,5,10$.
The surface plots of the relative motion transition probabilities $W_n^m$
for a fixed value of $d=10^{5}$, and for $r=10^{-6}, 10^{-5}, 10^{-4}$
are given in Figs.\ref{f:2wmnr6}, \ref{f:2wmnr5}, and \ref{f:2wmnr4},
respectively.  Fig.\ref{f:2wmnr6} shows that actually
there is no transitions to other states, if $rd=0.1$,
and $m,~n\le5$,
while for larger values of $m,~n$ some structure appears in the plot.
This structure becomes more complicated, when $rd=1$.
For $rd=10$, the picture is quite different from the adiabatic behavior,
although coefficient $r$ is still very small.

If $rd\gg m,n\sim 1$, then the leading term of the Laguerre polynomial
in (\ref{prob-gbigg}) yields
\be
W_n^m=\frac{(rd)^{m+n}}{m!n!} (1-r)^{d+1}\approx
\frac{(rd)^{m+n}}{m!n!} e^{-rd}.
\label{prob-rdbig}
\ee
Eq. (\ref{prob-rdbig})
clearly shows, that for moderate values of $m,n\ll rd$ (when the model
considered can be justified), {\it all\/}
transition probabilities are negligibly small.
This means that even
small time variations of the binding trap harmonic potential
result in the excitation of the states
with high values of the relative motion energy,
while the quantum state of the centre of mass motion is practically
unchanged under the same conditions.
This phenomenon can be interpreted as an indication of an
instability of the relative ion motion with respect to small perturbations
in the quantum regime.
We should note, however, that the quantitative
calculations of the transition probabilities to highly excited states of the
relative motion of real ions cannot be performed in the framework of
the singular oscillator potential model, since the condition (\ref{n-max})
is not fulfilled for these states.

\section{Conclusions}
The main results of the paper can be summarized as follows. We have
presented new asymptotical forms of the wave packets of the time
dependent singular quantum oscillator in the case of a very strong
repulsive centrifugal potential. We have shown that such a potential
can approximate to a certain extent the real Coulomb repulsion between two
ions in a trap.
We have found an extreme sensitivity of the transition amplitudes
between energy eigenstates to small variations of the coefficients of
the Hamiltonian.
It is worth noticing in this regard that the system of two trapped ions
interacting via the true Coulomb potential exhibits under certain
conditions a chaotic behaviour \cite{Blum,Nat,Moor}. Of course, one cannot
expect any chaos, in the strict sense of this word, in the framework of the
one-dimensional exactly solvable model. Nonetheless, our results might
be interpreted as an {\it indication of a possible\/} irregular
behaviour (which manifests itself just in a high sensitivity to small
changes of the initial conditions of parameters describing the system), if
one replaces the exactly solvable potential by the realistic one (which
cannot be treated analytically).

\newpage

\newpage

\newpage

\newpage

\newpage

\newpage

\newpage

\begin{thebibliography}{99}

\bibitem{Land} L. D. Landau and E. M. Lifshitz, {\it Quantum Mechanics\/}
(Addison-Wesley, Reading, 1965).

\bibitem{Camiz} P. Camiz, A. Gerardi, C. Marchioro, E. Presutti, and
E. Scacciatelli, J. Math. Phys. {\bf 12}, 2040 (1971).

\bibitem{DMM72} V. V. Dodonov, I. A. Malkin, and V. I. Man'ko,
    Phys. Lett. A {\bf 39}, 377 (1972).

\bibitem{DMM75}  V. V. Dodonov, I. A. Malkin, and V. I. Man'ko,
Int. J. Theor. Phys. {\bf 14}, 37 (1975);
 Teor. Mat. Fiz. {\bf 24}, 164 (1975)
    [Sov. Phys. - Theor.  \&  Math. Phys. {\bf 24}, 746 (1975)].

\bibitem{Inom} D. Peak and A. Inomata, J. Math. Phys. {\bf 10}, 1422 (1969).

\bibitem{Khan} D. C. Khandekar and S. V. Lawande,
J. Math. Phys. {\bf 16}, 384 (1975).

\bibitem{DMM74}  V. V. Dodonov, I. A. Malkin, and V. I. Man'ko,
    Physica {\bf 72}, 597 (1974).

\bibitem{gla63}
R.J. Glauber, Phys. Rev. Lett. {\bf 10}, 84 (1963).

\bibitem{DMM74b} V. V. Dodonov, I. A. Malkin, and V. I. Man'ko,
    Nuovo Cim. B {\bf 24}, 46 (1974).

\bibitem{Haut} A. Hautot, J. Math. Phys. {\bf 14}, 1320 (1973).

\bibitem{Niet} M. M. Nieto and L. M. Simmons, Jr., Phys. Rev. D
{\bf 20}, 1321, 1332, 1342 (1979).

\bibitem{CDM86}  S. M. Chumakov, V. V. Dodonov, and V. I. Man'ko,
    J. Phys. A {\bf 19}, 3229 (1986).

\bibitem{Prants} S. V. Prants, J. Phys. A {\bf 19}, 3457 (1986).

\bibitem{Datt} G. Dattoli, S. Solimeno, and A. Torre,
Phys. Rev. A {\bf 34}, 2646 (1986).

\bibitem{univ} V. V. Dodonov and V. I. Man'ko, in
    {\it Group Theoretical Methods in Physics, Proceedings of  the  Second
    International Seminar,
Zvenigorod, 24-26 November 1982}, edited by M. A. Markov, V. I. Man'ko,
and A. E. Shabad (Harwood Academic Publishers, Chur--London--New York, 1985),
    Vol. 1, p. 591.

\bibitem{Zhiv} V. V. Dodonov, V. I. Man'ko, and D. V. Zhivotchenko,
    Nuovo Cim. B {\bf 108}, 1349 (1993).

\bibitem{Ag}
G. S. Agarwal and S. Chaturvedi, J. Phys. A {\bf 28}, 5747 (1995).

\bibitem{Maam} M. Maamache, Phys. Rev. A {\bf 52}, 936 (1995);
    J. Phys. A {\bf 29}, 2833 (1996); Ann. Phys. (NY) {\bf 253}, 1 (1997).

\bibitem{Kaush} R. S. Kaushal and D. Parashar, Phys. Rev. A {\bf 55}, 2610
(1997).

\bibitem{Calog} F. Calogero, J. Math. Phys. {\bf 10}, 2191 (1969);
{\bf 12}, 419 (1971); B. Sutherland, J. Math. Phys. {\bf 12}, 246 (1971).

\bibitem{Hart} A. Hartmann, Theor. Chim. Acta (Berlin) {\bf 24}, 201 (1972).

\bibitem{167}
 V. V. Dodonov  and  V. I. Man'ko,
    in {\it Group Theory, Gravitation and Elementary Particle Physics\/},
    edited by A. A. Komar, Proceedings of Lebedev Physics Institute Vol.167
    (Nova Science, Commack, N.Y., 1987), p. 7.

\bibitem{sch35}
E. Schr\"{o}dinger, Naturwissenschaften, {\bf 23}, 844 (1935).

\bibitem{haroche}
S. Haroche, Nuovo Cim. B {\bf 110}, 545 (1995).

\bibitem{ys}
B. Yurke and D. Stoler, Phys. Rev. Lett., {\bf 57}, 13 (1986).

\bibitem{octavio}
O. Casta\~{n}os, R. J\'{a}uregui, R. L\'{o}pez--Pe\~{n}a,
J. Recamier, and V. I. Man'ko, Phys. Rev. A {\bf 55}, 1208 (1997).

\bibitem{yuen}
H. P. Yuen, Phys. Rev. A {\bf 13}, 2226 (1976).

\bibitem{walls}
D. F. Walls, Nature (London) {\bf 306}, 141 (1983).

\bibitem{schleich}
W. P. Schleich and J. A. Wheeler, Nature {\bf 326}, 574 (1987);
~J. Opt. Soc. Am. B {\bf 4}, 1715 (1987).

\bibitem{nikonov}
V. V. Dodonov, V. I. Man'ko and D. E. Nikonov, Phys. Rev. A {\bf 51},
3328 (1995).

\bibitem{hacyan}
S. Hacyan, Found. Phys. Lett. {\bf 9}, 225 (1996).

\bibitem{rosa} 
V. V. Dodonov, O. V. Man'ko, V. I. Man'ko, and L. Rosa, 
Phys. Lett. A {\bf 185}, 231 (1994).

\bibitem{paul}
W. Paul, Rev. Mod. Phys. {\bf 62}, 531 (1990).

\bibitem{vog1}
R.L. de Matos Filho and W. Vogel, Phys. Rev. Lett. {\bf 76}, 608 (1996).

\bibitem{gerry}
C. C. Gerry, Phys. Rev. A {\bf 55}, 2478 (1997).

\bibitem{wineland}
D.M. Meekhof, G. Monroe, B.E. King, W.M. Itano, and D.J. Wineland,
Phys. Rev. Lett. {\bf 76}, 1796 (1996).

\bibitem{glaconf}
R. J. Glauber, in  {\it Recent Developments in Quantum
Optics, Proceedings of the Intern. Conference on Quantum Optics 
(Hyderabad, India, January 1991)}, edited by R. Inguva
(Plenum Press, New York, 1993), p.~1.

\bibitem{schrama}
G. Schrade, V.I. Man'ko, W.P. Schleich, and R.J. Glauber,
Quantum Semiclass. Opt. {\bf 7}, 307 (1995).

\bibitem{sudphysscr}
V.I. Man'ko, G. Marmo, E.C.G. Sudarshan, and F. Zaccaria, 
in
{\it Proceedings of the Fourth Wigner Symposium (Guadalajara, Mexico,
July 1995)}, edited by N.~M.~Atakishiyev, T.~H.~Seligman, and K.-B.~Wolf
(World Scientific, Singapore, 1996), p.~421;
Phys. Scripta {\bf 55}, 528 (1997).

\bibitem{vog2}
R.L. de Matos Filho and W. Vogel, Phys. Rev. A {\bf 54}, 4560 (1996).

\bibitem{Blum} R. Bl\"umel, C. Kappler, W. Quint, and H. Walther,
Phys. Rev. A {\bf 40}, 808 (1989).

\bibitem{Comb} M. Combescure, Ann. Phys. (NY) {\bf 204}, 113 (1990).

\bibitem{Nat}  R. Bl\"umel {\it et al\/}, Nature {\bf 334}, 309 (1988).

\bibitem{Moor} M. Moore and R. Bl\"umel,
Phys. Rev. A {\bf 48}, 3082 (1993).

\bibitem{Ger89} C. C. Gerry and E. R. Vrscay, Phys. Rev. A {\bf 39}, 5717
 (1989).
\bibitem{Ger90} C. C. Gerry and T. Schneider, Phys. Rev. A {\bf 42}, 1033
 (1990).

\bibitem{Dod96} V. V. Dodonov, Phys. Lett. A {\bf 214}, 27 (1996).

\bibitem{Feng} M. Feng, K. Wang, J. Wu, and L. Shi, Phys. Lett. A
{\bf 230}, 51 (1997).

\bibitem{Bate} {\it Bateman Manuscript Project:
Higher Transcendental Functions}, edited by A.~Erd\'elyi
  (McGraw-Hill, New York, 1953).

\bibitem{BG} A. O. Barut and L. Girardello, Commun. Math. Phys.
{\bf 21}, 41 (1971).

\bibitem{Buz} V. Bu\v zek, J. Mod. Opt. {\bf 37}, 303 (1990).

\bibitem{Sat} G. Satya Prakash and G. S. Agarwal,
 Phys. Rev. A {\bf 50}, 4258 (1994).

\bibitem{Trif}
D. A. Trifonov, J. Math. Phys. {\bf 35}, 2297 (1994);
J. Phys. A {\bf 30}, 5941 (1997).

\bibitem{Brif}
C. Brif, Quant. Semiclas. Opt. {\bf 7}, 803 (1995);
C. Brif, A. Vourdas, and A. Mann, J. Phys. A {\bf 29}, 5873 (1996).

\bibitem{Onof} E. Onofri and M. Pauri, Lett. Nuovo Cim. {\bf 3}, 35 (1972).



\bibitem{MM70} I. A. Malkin and V. I. Man'ko, Phys. Lett. A {\bf 32},
243 (1970);
 I. A. Malkin, V. I. Man'ko, and D. A. Trifonov, Phys. Rev. D {\bf 2},
1371 (1970).

\bibitem{183}
  V. V. Dodonov and V. I. Man'ko, in {\em Invariants and the Evolution of
Nonstationary Quantum Systems}, edited by M. A. Markov,
Proceedings of Lebedev Physics Institute Vol. 183
(Nova Science, Commack, N. Y., 1989), p. 263.

\end{thebibliography}
\end{document}